%% file: main.tex
\documentclass[conference]{IEEEtran}
\IEEEoverridecommandlockouts
\usepackage{cite}
\usepackage{amsmath,amssymb,amsfonts}
\usepackage{graphicx}
\usepackage{textcomp}
\usepackage{xcolor}
\usepackage[a4paper, total={184mm,239mm}]{geometry}
\def\BibTeX{{\rm B\kern-.05em{\sc i\kern-.025em b}\kern-.08em
    T\kern-.1667em\lower.7ex\hbox{E}\kern-.125emX}}
    
\usepackage{amsmath,bm}
\usepackage{amssymb}
\usepackage{amsthm}
\usepackage{mathrsfs}
\usepackage{relsize}
\usepackage{amsfonts}
\usepackage{cases}
\usepackage{mathtools}
\usepackage{pifont}
\usepackage[T1]{fontenc}
\usepackage{algorithm}
\usepackage{algpseudocodex}
\usepackage{booktabs}
\usepackage{enumitem}

\usepackage{hyperref}
\usepackage{cleveref}
\hypersetup{
            colorlinks=true,
            linkcolor=blue,
            anchorcolor=blue,
            citecolor=blue}

\usepackage{multirow}
\ifCLASSOPTIONcompsoc
    \usepackage[caption=false,font=normalsize,labelfont=sf,textfont=sf]{subfig}
\else
    \usepackage[caption=false,font=footnotesize]{subfig}
\fi
\usepackage{xcolor}
\usepackage[perpage,para, stable]{footmisc}
\usepackage{ulem}
\usepackage[para,online,flushleft]{threeparttable}

\DeclareMathOperator*{\round}{round}
\DeclareMathOperator*{\clamp}{clamp}

\begin{document}
\bstctlcite{IEEEexample:BSTcontrol}


\title{Algorithm-hardware co-design for Energy-Efficient A/D conversion in ReRAM-based accelerators
\thanks{This work is supported by Key-Area Research and Development Program of Guangdong Province (2021B0101310002), NSF China(62032001), 111 Project (B18001)}
}

\author{
    \IEEEauthorblockN{Chenguang Zhang$^{1,2}$, Zhihang Yuan$^{1,2}$, 
    Xingchen Li$^{1,2}$, 
    Guangyu Sun$^{1,3,*}$\thanks{*Corresponding author.}}
    \IEEEauthorblockA{$^1$ School of Integrated Circuits, Peking University, Beijing, China}
    \IEEEauthorblockA{$^2$ School of Computer Science, Peking University, Beijing, China}
    \IEEEauthorblockA{$^3$ Beijing Advanced Innovation Center for Integrated Circuits, Beijing, China}
    \IEEEauthorblockA{zhangchg@stu.pku.edu.cn, \{yuanzhihang, lixingchen, gsun\}@pku.edu.cn}
}

\maketitle
\input{tex/0_abstract}
\input{tex/1_introduction}
\input{tex/2_background}

\input{tex/4_1_method}
\input{tex/4_2_hw_design}
\input{tex/4_3_codesign}
\input{tex/5_experiment}

\input{tex/6_conclusion}

\bibliographystyle{IEEEtran}
\bibliography{./bib/IEEEabrv,./bib/ACMabrv,./ref}

\end{document}

%% file: tex/0_abstract.tex
\begin{abstract}

Deep neural networks are widely deployed in many fields.
Due to the in-situ computation (known as processing in memory) capacity of the Resistive Random Access Memory (ReRAM) crossbar, 
ReRAM-based accelerator shows potential in accelerating DNN with low power and high performance.
However, despite power advantage, such kind of accelerators suffer from the high power consumption of peripheral circuits, especially Analog-to-Digital Converter (ADC), 
which account for over 60 percent of total power consumption. This problem hinders the ReRAM-based accelerator to achieve higher efficiency.

Some redundant Analog-to-Digital conversion operations have no contribution to maintaining inference accuracy, 
and such operations can be eliminated by modifying the ADC searching logic.
Based on such observations, we propose an algorithm-hardware co-design method and explore the co-design approach in both hardware design and quantization algorithms. 
Firstly, we focus on the distribution output along the crossbar's bit-lines 
and identify the fine-grained redundant ADC sampling bits. 
To further compress ADC bits, we propose a hardware-friendly quantization method and coding scheme, 
in which different quantization strategy was applied to the partial results in different intervals. 
To support the two features above, we propose a lightweight architectural design based on SAR-ADC\@. 
It's worth mentioning that our method is not only more energy efficient but also retains the flexibility of the algorithm. 
Experiments demonstrate that our method can reduce about $1.6 \sim 2.3 \times$ ADC power reduction.

\end{abstract}

%% file: tex/1_introduction.tex
\section{Introduction}

With the wide adoption of deep neural networks (DNNs), an increasing number of applications utilize multiple DNNs to achieve state-of-the-art performance. 
The general and energy-efficient DNN 
inference acceleration demands surge at the edge such as automotive, robotics, and IoT devices.
Among various accelerator architectures~\cite{FPGAMuitiTask, Boroumand2021-mensa}, 
Processing-In-Memory (PIM) can avoid frequent data movement between memory and computing units 
showing great potential in breaking the bandwidth limitation and enhancing power efficiency. 
Additionally, the ReRAM-based analog computation~\cite{shafiee2016isaac} provides further advantages to PIM
due to its features of nonvolatile, high density, and in-situ matrix-vector-multiplications (MVMs)\@.

However, the overall energy efficiency of ReRAM-based PIM accelerators is impaired by the high overhead of analog-to-digital converters (ADCs) that digitize the analog results from crossbars (XBs).
High-resolution ADC even consumes much more power and area than the ReRAM crossbar itself\cite{Li22Tailor}. 
%
To eliminate the ADC bottleneck, existing solutions can be broadly classified into two types: 
algorithm-level techniques exploit weight sparsity and reduce the computation amount to reduce ADC usage~\cite{qu2021asbp, zhu2019configurable}. 
However, annoying retraining limits the range of applicable models at deployment.
Architecture-level techniques replace high-resolution ADC with other highly customized circuits~\cite{chou19cascade, chi2016prime} or ADC designs~\cite{sun2020energy}. 
These adaptations are not so feasible in practice due to the consideration of manufacturing and analog circuit design complexity.

Targeting to improve energy efficiency while retaining the flexibility of the algorithm, 
in this paper, we adopt an approach at different levels: 
we find that a large proportion of ADC output bits are redundant during each A/D conversion process (Section~\ref{motivation}), 
and contribute little to the inference accuracy, due to the imbalanced distribution and inherent fault tolerance of DNNs. 
Based on this observation, we propose a configurable coding scheme
to avoid redundancy generation and thereby improve energy efficiency, 
and implement the coding-decoding by a modified successive approximation register (SAR) ADC design.
Our key contributions are:
\begin{itemize}[leftmargin=*]
\item We analyze the distribution of analog values from crossbar bit-lines and quantify the redundancy in ADC output coding (Section~\ref{analysis}).
\item We proposed a hardware-friendly quantization method, Twin range quantization, that can flexibly adjust quantization intervals based on the value distribution, and an efficient coding scheme for ADC outputs (Section~\ref{Methods}).
\item We customize the SAR ADC control logic to support configurable quantization levels without changing analog circuits (Section~\ref{hw-design}).
\item An algorithm-hardware co-optimization scheme is proposed to search for the optimal parameters for each DNN layer, maximizing energy saving while minimizing accuracy loss (Section~\ref{codesign}).
\end{itemize}
Our techniques require \textit{no DNN retraining} and \textit{no customization of ADC's analog part}. 
The proposed ADC modification is transparent to DNN models and 
is orthogonal to any other approaches based on model compression and auto-ML methods. 


%% file: tex/2_background.tex
\section{Background and Motivation}\label{background}

\subsection{Analog PIM basic}

\paragraph*{\textbf{PIM macro}}
An analog PIM macro is the basic unit of a ReRAM-based accelerator,
consisting of a crossbar array and peripheral circuits.
The ReRAM crossbar functions as an MVM engine. 
The voltages applied to the word lines (WLs) multiply 
with the conductance of the cells. 
Currents of each cell accumulate along the bit lines (BLs), 
producing analog the MVM result, i.e., $\mathbf{I}_i=\sum_j \mathbf{G}_{i,j}\mathbf{V}_j$. 
ADC converts the currents at each BL to digital outputs.

\begin{figure}[tbhp]
    \centering
    \includegraphics[width=0.94\linewidth]{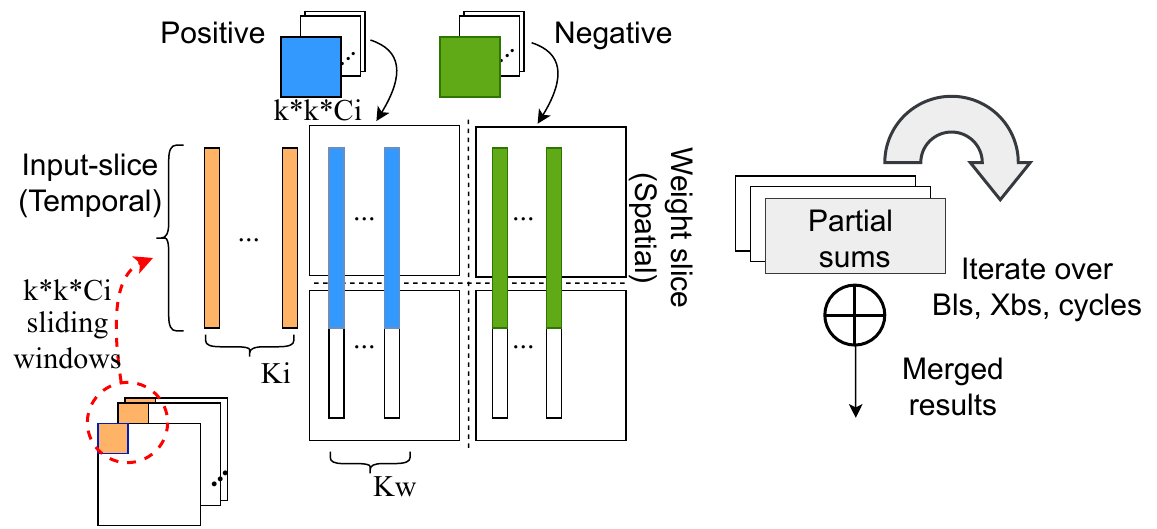}
    \caption{Mapping convolutional layers. $K_w, K_i$ are the bit-width of the weight and input activation respectively.}%
    \label{fig::base_mapping}
    \vspace{-0.5cm}
\end{figure}

\paragraph*{\textbf{Map Neural Networks to Accelerator}} 
To map a neural network to the PIM macro,
elements in input features are fed to DACs,
and weights are stored as the conductance of the ReRAM cells.
As shown in Fig.~\ref{fig::base_mapping}, 
a convolutional kernel (with $k \times k$ kernels, $C_i$ input channels) is transformed to MVM in different sliding windows~\cite{shafiee2016isaac}. 
Resolution limitation of DAC and ReRAM cells~\cite{IRDS21MM} necessitate bitwise sliced mapping:
a weight value is partitioned and mapped to cells on different BLs,
input vectors are fed to DAC as bit slice cycle by cycle. 
When not fitting into a crossbar pair, a layer is partitioned and mapped to multiple crossbars. 
This sliced \textbf {intermediate} MVM results along BLs, cycle, and crossbars
are merged by shift-and-add and accumulator in the digital domain, 
getting the final MVM result,
which is sent to other functional units or crossbars for further processing.

\subsection{Quantization in Datapath} 
Quantization can effectively compress neural networks. 
The uniform quantization projects a real number $x$ to a $k$-bit integer value $x_q$, 
by dividing a given range of real values into several partitions into $2^k-1$ $\Delta_x$-sized intervals,
and rounding ($\text{round}(\cdot)$) the value to the nearest integer from $0$ to $2^k-1$, as shown in Eq.~\ref{sym_quant}.
\begin{equation}
    \begin{aligned}
        x_q  =  Q_k(x,\Delta)=\Delta_x \clamp(\round(\frac{x}{\Delta_x}),0,2^{k}-1), \\
        \text{where}, \Delta_x = {(b-a)}/{(2^{k}-1)}%
    \end{aligned}
    \label{sym_quant}
\end{equation}
Low bit-width helps in reducing power, area, and latency. 
Within a ReRAM-based accelerator, quantization not only happens at the algorithm level. 
ADC with lower-than-ideal resolution also behaves as a quantizer, introducing extra quantization errors to the intermediate MVM results.
This behavior was not the designer's intent and needed to be specifically considered in the DNN quantization algorithm design.


\subsection{ADC Preliminaries and Previous Works}%
\label{ADC_preliminaries}
Lossless conversion requires ADC with a resolution no less than $R_{ADC, ideal}$, i.e.,
\begin{align}
    &R_{ADC, ideal} &=& \log_2(S) + R_{DA} + R_{cell} + \delta,\\
    &\delta &=& 0\; \text{if } R_{DA}\ge 1\text{ and }R_{cell} \ge 1\text{ else }-1, \notag
\end{align} 
where $R_{cell}, R_{DA}, S$ are the number of bits a cell represents, the resolution of DAC, and the size of the crossbar, respectively. 
The overall ADC energy is formulated as:
\begin{align}
    E_{ADC, tot} =&
    \overbrace{\text{len (input)} \times \text{size}_{w}}^{\# \text{ of MVMs / inference}} 
    \times \overbrace{\frac{K_w}{R_{cell}} \times \frac{K_i}{R_{DA}}}^{\# \text{ of A/D conversions / MVM}} \\
    \times &E_{convert},
\end{align}
where $E_{convert}$ is the energy per A/D conversion. 
As the power of ADC scales exponentially with ADC resolution $R_{ADC}$ high-resolution ADCs are not competitive in terms of power and area. 
To overcome this challenge, 
solutions proposed by recent works can be categorized into three orthogonal levels: 
circuit, architecture, mapping, and algorithm.
\begin{itemize}[leftmargin=*]
    \item At the algorithm level,
          low-bit quantization~\cite{qu2021asbp, zhu2019configurable, sun2020energy} is widely used to reduce   the bit-width of the weight and input activation ($K_w, K_i$);
          and pruning techniques~\cite{ma2020tiny, Shin2022Effective, qu2021asbp} are used to reduce
          required MVMs per inference.
    \item At the mapping stage,
          weight encoding is introduced to obtain smaller conductance values ($K_w$)~\cite{Andrulis2023RAELLA, shafiee2016isaac};
          weight slicing and input slicing is explored in {RAELLA}~\cite{Andrulis2023RAELLA}
          to trade off ADC resolution (i.e., $E_{convert}$) and the number of A/D conversions per MVM\@.
    \item At the architecture level~\cite{shafiee2016isaac, liu202033}, 
          1-bit DAC and 1-bit ReRAM cells ($R_{DA}, R_{cell} = 1$) are widely adopted,
          and $R_{DAC, ideal} = \log_2(S) + 1$.
          Multi-bit input fed to DAC as bit series, aggregated by shift-and-add operations.
    \item At the circuit level, to reduce $E_{convert}$,
          ADCs are customized with a non-uniform quantization scheme~\cite{sun2020energy};
          or discard, with all the operations conducted in the analog domain.
\end{itemize}

\subsection{A/D operations and coding scheme}
\begin{figure}
    \centering
    \includegraphics[width=0.98\linewidth]{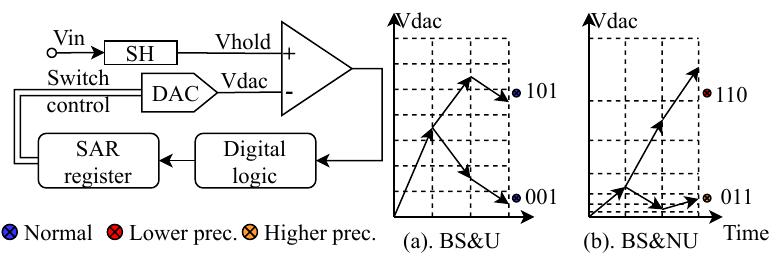}
    \caption{Conventional SAR ADC with (a) uniform and (b) non-uniform grid}%
    \vspace{-0.7cm}
    \label{fig::sar_adc}
\end{figure}
SAR ADC is the most suitable ADC for analog PIM macro, 
which is more energy-efficient than other ADCs in the target resolution and frequency~\cite{murmann2012energy},
and strongly benefits from process technology scaling~\cite{murmann2013energy}.

The digital-to-analog converter (DAC) generates predefined threshold voltages $V_{th, idx(k)}$ for comparison with the input analog signal ($V_{hold}$). 
SAR logic generates digital signals ($idx(k), k=1,\ldots, K$) to guide a Binary Search (BS) to obtain a $K$-bit output code ($idx(K+1)$),  representing the index of the closest voltage level.
As shown in Fig.~\ref{fig::sar_adc}, the horizontal dash grid represents 
reference voltages, 
points ($\otimes$) represent the sampled values to be converted, 
and arrows consist of searching traces. 
$(\cdot)_2$ represents of binary code of the index tested at each step. 
Starting from $(10\ldots0)_2$, at the $k$th step, SAR logic always tries the ${(K\!-\!k)}$th bit with ``1'', 
and then replaces the ``1'' by the comparator's output ($D_{k}$) in the next step.
In $K$ cycles, The binary code is iteratively filled from MSB to LSB\@, 
and getting $(D_{1}\ldots D_{K-1})_2$ at the end. 
\begin{equation}
    idx(k) = \begin{cases}
             (10,\ldots,0)_{2}, &k=0\\
             (\underbrace{ D_{1} \cdots D_{k-1}}_{k-1} 1 \cdots 0)_{2}, & k=1, \ldots, k-1
             \end{cases}
    \label{eq::sar_logic}
\end{equation}
We refer to each step as an \textit{A/D operation}, and full progress as an \textit{A/D conversion}.
For a SAR ADC with $K$-bit resolution, energy consumption is proportional to the number of A/D operations,
\begin{equation}
    E_{convert} = e_{\text{op}} N_{A/D\_ops}
\end{equation}
where $e_{\text{op}}$, and $N_{A/D\_ops, i}$ represent energy per A/D operation, 
and the number of A/D operations required by each conversion, respectively. 

For conventional Uniform (U) ADCs,
threshold voltages are equally spaced ($(k-\frac{1}{2}) \cdot \text{LSB}, k=1,\ldots, 2^K-1$).
In contrast, Non-uniform ADC (NU) performs BS on a customized grid,
which has a higher density in the range with more values, as shown in Fig.~\ref{fig::sar_adc}~(b). 
With lower $R_{ADC}$, \uline{NU ADC achieves similar accuracy as U ADC, 
but with a large reduction in $e_{op}$}, also lower ADC power consumption.
\uline{Both of them have fixed $N_{A/D\_ops}$ for each conversion.}

\subsection{Motivation}\label{motivation}

Customizing ADCs and retraining DNN models restrict the algorithm flexibility of the accelerator.
Therefore, we avoid modifying the analog characteristics of ADC\@.
From the above dive into the ADC and related works,
we notice that the factor, $N_{A/D\_ops}$, 
which is determined by the searching process (i.e., coding) of SAR logic,
is an indicator of A/D conversion efficiency. 
BS is a good choice for numerical comparison in general 
but may be not the optimal choice for a given distribution.
This motivates us to \uline{explore the optimal coding scheme (with least $N_{A/D\_ops}$)
and configurable SAR logic to achieve adaptive energy-efficient A/D conversions, 
which is orthogonal to the above techniques}.

%% file: tex/4_1_method.tex
\section{Twin Range quantization}\label{Methods}
In this section, we first identify the redundancy in the output coding scheme, 
and then we introduce our Twin Range quantization and coding scheme to remove such redundancy.
Finally, we present the implementation of configurable ADC coding/decoding. 

\subsection{Value distribution at BLs and redundancy}\label{analysis}
\begin{figure}[tbp]
    \centering
    \subfloat[]{
        \includegraphics[width=0.36\linewidth]{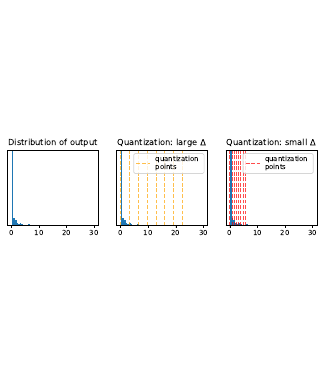}%
        \label{distribution}
    }
    \subfloat[]{
        \includegraphics[width=0.54\linewidth]{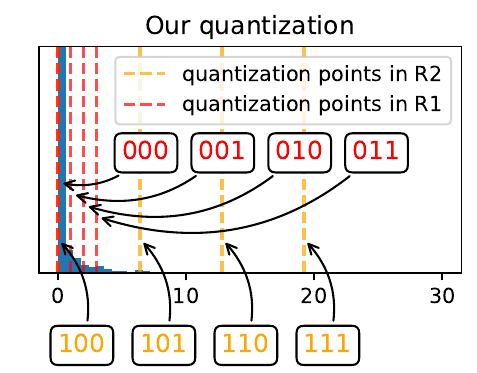}%
        \label{twin_ranges}
    }
    \caption{(a) Distribution of the output of crossbar's BLs. (b) Twin ranges quantization. }%
    \vspace{-0.5cm} 
\end{figure}
While it is commonly assumed that the activation in neural networks follows a Gaussian distribution~\cite{Low_bit_quant_iccvw2019}, 
the value distribution of the ReRAM crossbar's BLs has not been thoroughly investigated. 
Fig.~\ref{distribution} illustrates the distribution of the BLs, 
revealing a highly imbalanced distribution where the majority of samples 
are concentrated in a small interval close to zero. 

Generally, achieving lossless data bit width compression on such a skew distribution is challenging:
we need to retain both the \textit{numerical range} of large values while minimizing the \textit{quantization error} of small values.
Huffman coding, which is a classical idea of using variable-length codes 
to improve the coding efficiency of skewed data, 
inspires us to explore more efficient intermediate coding schemes used in A/D conversion. 
Additionally, the inherent fault tolerance of DNN 
enlightens us to reduce the precision of values that are not sensitive to inference accuracy.
To this end, we identify samples of two kinds in the skewed distribution:
(\textbf{T1}) the majority that concentrate in a narrow range (R1) and (\textbf{T2}) the minority that scatter in a wide range (R2), 
and correspondingly, two compression strategies.
\subsection{Quantization abstraction}\label{quant}
\begin{figure}[tbp]
    \centering
    \subfloat[] {
        \includegraphics[width=0.33\linewidth]{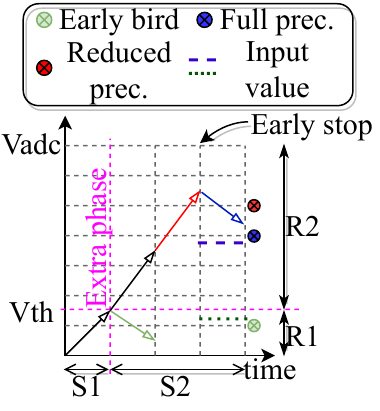}%
        \label{fig:transform_curve}
    }
    \subfloat[]{
        \includegraphics[width=0.55\linewidth]{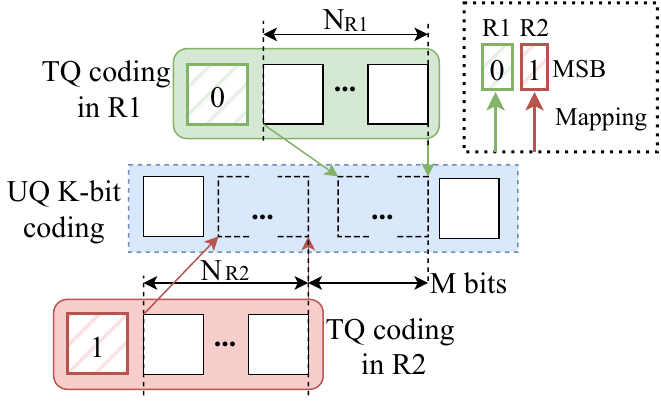}%
        \label{fig:bit_codding}
    }
    \caption{(a) DAC output with two searching strategies. (b) The bit mapping of the ADC output code.}%
    \vspace{-0.5cm} 
\end{figure}
From the perspective of A/D conversion,
we can apply different SAR logic to the two ranges for less A/D operations:
\begin{itemize}[leftmargin=*]
    \item ``Early birds'': 
    With the expense of a 1-bit comparison for each conversion,
    we can apply a biased search: 
    if R1 is small and dense enough (``sweet spot''),
    most conversions can be completed with fewer A/D operations, 
    but without precision loss. 
    As shown in Fig~\ref{fig:transform_curve},
    ``early bird'' (\textcolor[HTML]{82B366}{green}) in R1 is approximated just in $1+1$ step.
    \item ``Early stopping'': 
    If T2 is sparse and not sensitive to inference accuracy,
    we can stop the conversion progress even though the output code is not fully determined,
    resulting in a reduced sensing precision but retaining the numerical range. 
    In figure~\ref{fig:transform_curve}, ``early stopping'' is made (\textcolor[HTML]{FF3333}{red}) after two steps before 
    a full precision (\textcolor[HTML]{3333FF}{blue}) conversion completed. 
\end{itemize}
Two strategies play different roles in balancing power and computation accuracy: 
``Early birds'' earns the $N_{A/D\_ops}$ without precision degradation, 
but is required to find the ``sweet spot''. 
``Early stopping'', has to determine whether to force a stopping or continue with full precision.
Therefore, systematic analysis and calibration are required to maximize the gain in power and latency (Section~\ref{codesign}). 


From the perspective of the quantization algorithm,
the above two strategies can be abstracted as a twin-range quantization (\textbf{TRQ}), 
a narrow range R1 to precisely quantize the small dense values, and a wider range R2 to cover the large sparse values;
each range is quantized uniformly but is assigned a distinct scaling factor. 
Different from previous works, this non-uniform stems from the biased searching strategy, 
resulting in grids that align with the full precision searching grids,
as seen in Fig.~\ref{twin_ranges} (red dashed vertical grid for R1 and orange for R2).
And the quantization function $\text{T}_k$ can be formulated as Eq~\ref{non_uniform_quant}:
\begin{IEEEeqnarray}{rCl}
    &\text{T}_k(x, \Delta_{\text{R1}}, \Delta_{\text{R2}}, N_{R1}, N_{R2})=\begin{cases} Q_{k-1}(x,\Delta_{\text{R1}}), x\leq \theta \\ Q_{k-1}(x,\Delta_{\text{R2}}), otherwise, \end{cases} \IEEEnonumber \\
    &\theta = 2^{N_{R1}}\Delta_{\text{R1}} , \text{R1}=\left [0, \theta \right ) , \text{R2}=\left [\Delta_{\text{R2}},+\infty \right ), \label{non_uniform_quant}
\end{IEEEeqnarray}
where $Q_k(x,\Delta)$ performs $k$ bit uniform quantization, with the scaling factor $\Delta$ (mentioned in Eq.~\ref{sym_quant}).
$N_{R1}, N_{R2}$ and $\Delta_{\text{R1}}, \Delta_{\text{R2}}$ are quantization bits and scaling factors of the two ranges. 
\textbf{Note} that, 
as a type of post-training quantization scheme (PTQ)~\cite{PTQ_4bit_rapid_deployment_nips2019}, the parameters can be easily calibrated to adapt to different DNNs by our algorithm-hardware co-design method (in Section~\ref{codesign}), \textbf{no retraining is required!}
Our quantizer is the \textit{behavior abstraction of A/D conversion of SAR-ADC at BLs} and is orthogonal to the other quantization on W/A/I. 
%
\subsection{Coding scheme}\label{sec::coding_scheme}
We devise a bitmap to encode the values quantized by our TRQ. As shown in Fig.~\ref{fig:bit_codding}, 
The most significant bit (MSB) indicates which range a value belongs to, ``0'' for R1 and ``1'' for R2. 
The remaining $N_{Rx}$ bit(s) is an unsigned uniform coding for the approximated value in each range. 
To make the quantization grid align with the full precision grid, we make  
\begin{equation}
    \Delta_{\text{R2}}=2^M\Delta_{\text{R1}}, 
    \label{eq::range_constrain}
\end{equation}
where $m$ is an unsigned integer parameter. 
In the decoding stage, the output code with MSB=1 is shifted left by $m$ bits to align with the value from R1. 

As can be seen, \textbf{TRQ requires neither codebook nor analog DAC modification},
which greatly simplifies the hardware design, which will be discussed in Section~\ref{hw-design}.


%% file: tex/4_2_hw_design.tex
\subsection{Hardware design}%
\label{hw-design}

\subsubsection{Overall architecture}
\begin{figure}[tb]
    \centering
    \includegraphics[width=0.99\linewidth]{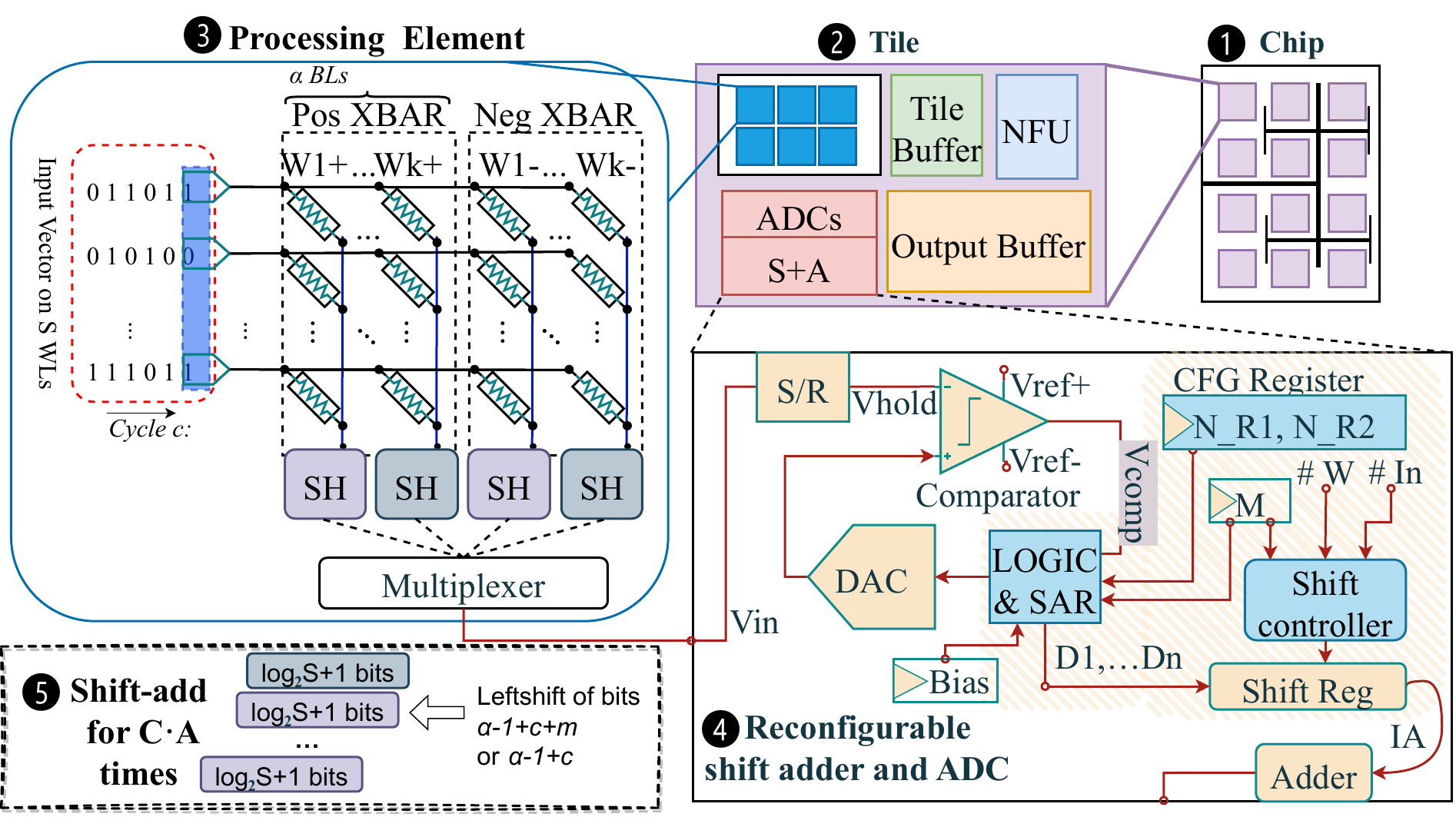}
    \caption{Overall architecture}
    \label{fig:ARCH}
	\vspace{-0.7cm} 
\end{figure}

In this paper, we follow the overall architecture of ISAAC~\cite{shafiee2016isaac}, shown in Fig.~\ref{fig:ARCH}. 
The accelerator (\ding{202}) consists of multiple tiles connected by a global bus. 
Each tile (\ding{203}) contains Neural Function Units (NFU), PE array, ADCs, Shift and Add (S+A) module, input buffer, and output buffer. 
The PE (\ding{204}) is used for accelerating the MVM operation. 
The accumulated product current at the BL end is converted to voltage values by the trans-impedance amplifiers (TIAs) and retained by the Sample-and-Hold (SH) circuit.
We implement lightweight modifications to the SAR logic (\ding{205}) and the S+A module (\ding{206}). 
Here, the voltage vectors are initially encoded into compact digital codes by ADCs 
and subsequently decoded and accumulated by the S+A module in a column-wise and cyclical manner. 
ADCs and S+A modules operate in a time-division manner, shared by the PEs.
The output feature maps are stored in the output buffer before being routed to other tiles or off-chip memory.
As both the \textit{coding} and \textit{decoding} steps are handled by hardware, 
no additional \textit{software overhead} is required.
The \textit{analog components} of the ADC remain unchanged, and the \textit{original resolutions} of the SAR ADC are preserved.


\subsubsection{Pipeline with Configurable Resolution}

\paragraph{\textbf{SAR logic with twin ranges}}
We apply TRQ, a customized searching strategy, as illustrated in Fig.~\ref{fig:transform_curve}.
First, the approximation logic checks whether the sampled voltage is located in R1 
i.e., $\left [0,2^{N_{R1}}\Delta_{\text{R1}} \right )$, 
with an \textit{extra detection phase}, 
%
As illustrated in Fig.~\ref{fig:transform_curve}, BS is performed in R1 and R2, with respective step sizes of 
$\Delta_{R1}$ and $\Delta_{R2}$. 
Both R1 and R2 are exponential multiples of 2, 
with $V_{grid}=V_{ref}/R_{ADC}$ representing the minimum voltage step, 
which can be configured by adjusting $V_{ref}$ of ADC or gain of the TIA amplifier. 
We achieve this by simply adding extra approximation logic to SAR control.

\paragraph{\textbf{Shift and Add module}}
The modified ADC generates a compact digital code in the format mentioned in Section~\ref{sec::coding_scheme}, 
which can not be accumulated directly by an arithmetic adder.
The radixes of the output code with MSB of 1 is $2^M$ times the output code with MSB equal to 0, 
which requires decoding before arithmetic operations.
Thanks to the hardware-friendly encoding scheme, decoding requires only shift operations. 
We add an extra shift control to the existing S+A module to support the above decoding:
Determining the MSB, 
before adding to the partial sum,
the ADC's output code is shifted left by $M$ bits (\ding{206}) (MSB is 1) or not (MSB is 0),
and the MSB is discarded.

\paragraph{\textbf{Configurable register}}
The configuration, including ADC output bit-width ($N_{R1}, N_{R2}$), step size ($\Delta_{R1}, \Delta_{R2}$), non-uniform degree ($M$), 
an offset of $R1$ ($Bias$, used to adapt outlier conditions, as discussed in Section~\ref{codesign})
are restored in the resister near the ADC and the Shift and Add module. 
The sensing precision can be configured as any bits below ADC resolution ($R_{ADC}$), 
and the non-uniform can be configured from 0 to $R_{ADC} - N_{R2}$.
Such flexibility enables our design can adapt to various DNNs and be compatible with other algorithm-level compression techniques.
Besides, our ADC design can be configured as either twin ranges mode or U ADC mode. 



%% file: tex/4_3_codesign.tex
\section{Algorithm and Hardware Co-optimization}\label{codesign}
Section~\ref{Methods} introduces our TRQ algorithm and its corresponding hardware implementation. 
Our design is capable of compressing redundant A/D operations, 
thereby reducing ADC energy consumption and improving power efficiency.
In this section, we propose a parameter-searching algorithm to determine the configuration with 
the maximum energy reduction while meeting accuracy constraints.

\subsection{Parameter Calibration}%
\label{subsec::para_calib}

To pinpoint the optimal values for R1 and establish an ``early stopping'' threshold for R2, 
we employ a layer-by-layer parameter search as detailed in Algorithm~\ref{algo::search}. 
For each layer, the algorithm first determines the distribution type of the layer's BL output ($y$) 
and then identifies the best setting for that layer. 
We sample $C$ candidates of $V_{grid}$ uniformly from the interval 
$[\alpha \frac{y_{\max}}{2^{R_{ADC}}-1}, \beta \frac{y_{\max}}{2^{R_{ADC}}-1}]$,
with $\alpha$ and $\beta$ defining the search range. 
We try to minimize the energy consumption:
\begin{equation}
    \begin{aligned}
    &(N_{R1} \;\text{or}\; M)= \mathop{\arg\min}_{(N_{R1} \;\text{or}\; M)} e_{\text{op}}(N\cdot\nu + \sum^{N_D}_{i=1} N_{A/D\_ops,i}),  \\
    &N_{A/D\_ops,i} = \begin{dcases}
        N_{R1}, & i \in R1 \\
        N_{R2}, & else,    %
    \end{dcases} \; \nu = \begin{dcases}
        1, & bias = 0 \\
        2, & else.    %
    \end{dcases}
    \end{aligned}
    \label{Eq:E-w.r.t-M}
\end{equation}

Here, $N_D$, $e_{\text{op}}$, and $N_{A/D\_ops, i}$ represent the number of samples, energy per A/D operation, 
and the number of A/D operations required by each sample, respectively. 
The item $N \cdot \nu$ represents the overhead of the pre-detection phase.
For each parameter combination, the quantization MSE is evaluated as
\begin{equation}
    \min_{\Delta_{\text{R2}}} \text{MSE}(\text{T}_k(x, \Delta_{\text{R1}}, \Delta_{\text{R2}}, N_{R1}, N_{R2}),y).
    \label{search_target}
\end{equation}
to find the optimal $V_{grid}$. 
End-to-end accuracy is checked after optimization of all layers, 
the search process iterates over descending $N_{R2}$ until the accuracy drops below the threshold $\theta$.
Finally, the resulting twin range quantization is compared with standard uniform quantization to select the best approach for each layer.
%
\begin{algorithm}
    \renewcommand{\algorithmicrequire}{\textbf{Input:}}
    \renewcommand{\algorithmicensure}{\textbf{Output:}}
    \caption{Parameter Searching}
    \begin{algorithmic}[1]
    \Require A pre-tuned quantized DNN model, prediction accuracy threshold $\theta$ 
    \Ensure $M, R1$ 
    \State $N_{\max}  \gets R_{ADC} - 1$ 
    \State $Acc \gets$ to evaluate the inference accuracy of a well-tuned quantized model 
    \While{True}
        \For {$l = 1 \rightarrow N$} 
            \State Judge the distribution types $T$ of current layer 
            \For {$V_{grid} \in range(\alpha \frac{y_{\max}}{2^{R_{ADC}}-1}, \beta \frac{y_{\max}}{2^{R_{ADC}}-1})$} 
                \State $R_{ideal} = ceil(\log2(y_{l,\max}-y_{l,\min}+1))$
                \State $N_{R2} = \min(N_{\max}, R_{ideal})$
                \If{$T$ is $\text{case}_{ideal}$ or $\text{case}_{N}$}
                    \State search for $N_{R1}$ and $bias$ minimize Eq.~\ref{Eq:E-w.r.t-M} 
                    \State set $M, \Delta_{\text{R1}}$ refer to Eq.~\ref{eq::dense_constrain} 
                    \Else 
                    \State set $N_{R1} = N_{R2}$ 
                    \State search for $M$ and $bias$ minimize Eq.~\ref{Eq:E-w.r.t-M} 
                    \State set $\Delta_{\text{R1}} = 2^{R_{ideal}-N_{R2}-M}$ 
                \EndIf 
                \State set $\Delta_{\text{R2}}$ (Eq.~\ref{eq::range_constrain}), record best loss (Eq.~\ref{search_target}) 
            \EndFor 
            \State Find optimal $V_{grid}$ as Eq.~\ref{search_target} 

        \EndFor 

        \State $Acc' \gets$ to evaluate the inference accuracy 
        \If{$Acc-Acc'>\theta$}
            \State Break 
        \Else 
            \State $N_{\max} \gets N_{\max} -1$ 
        \EndIf 
        \State Compare with uniform quantization with $N_{R2}$ bits 
    \EndWhile 
    
    \end{algorithmic}%
    \label{algo::search}
  \end{algorithm}
%
%
$N_{R1}$ is deduced regarding the distribution type of the layer's BL output.
For the \textit{ideal cases} in Section~\ref{quant}, 
which performs lossless A/D conversion in $R1$, 
We can deduce,
\begin{equation}
    \Delta_{\text{R1}} = 1, N_{R2} + M = R_{ideal}, bias=0, \nu = 1 \label{eq::dense_constrain}.
\end{equation}

\subsection{Compatibility for various types of distribution}\label{subsec::distribution} 
The normal-like distribution with strong unimodality (i.e., low-variance), is also treated as the ideal case, 
except for $\text{Offset} = bias \cdot V_{ref}/2^{M}$ 
is introduced as an extra threshold to identify the range $R1$ and 
$bias$, which is searched over the integer from $0$ to $2^{M}-1$,
is concatenated to the left side of the coding from $R1$ in the decoding progress. 
In other cases, such as weak unimodal, multi-modal, and flattened distribution, 
if a ``sweet spot'', $R1$ may not be found, 
``early stopping'' strategy is performed in both ranges. 
To reduce the search space, $N_{R1}$ and $N_{R2}$ set to the same values, 
and both $\Delta_{\text{R1}}$ and $\Delta_{\text{R1}}$ are searched to minimize quantization error Eq.~\ref{search_target}.

%% file: tex/5_experiment.tex
\section{Evaluation}\label{Evaluation}

\subsection{Experiment Settings}
\begin{figure*}[t]
  \centering
  \subfloat[]{
  \includegraphics[width=0.33\linewidth]{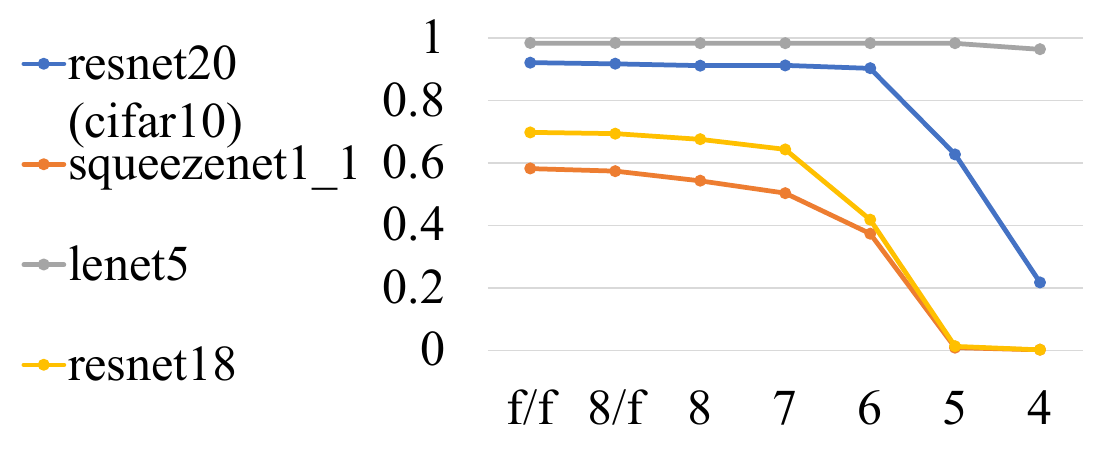}
  }
  \subfloat[]{
    \includegraphics[width=0.23\linewidth]{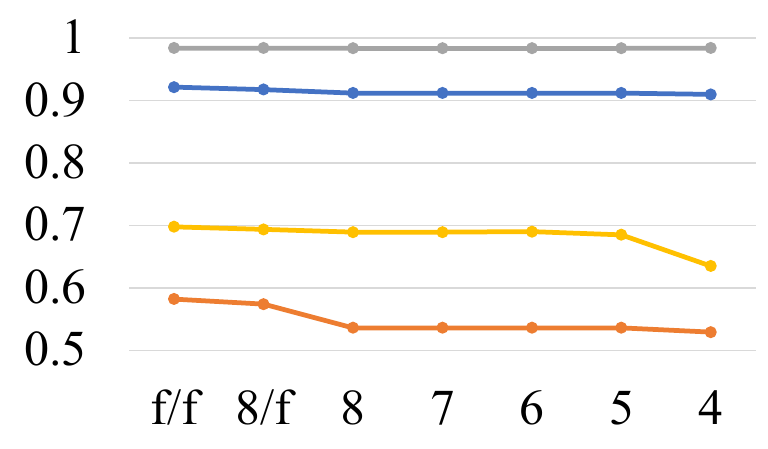}
  }
  \subfloat[]{
    \includegraphics[width=0.23\linewidth]{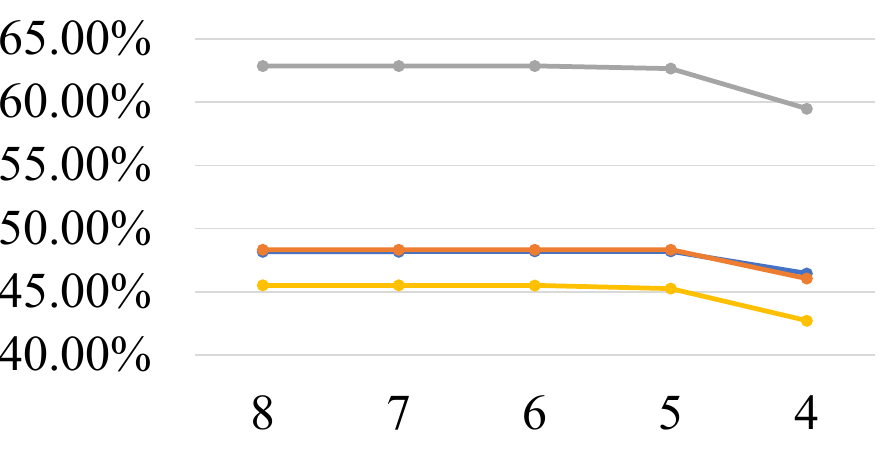}
  }
  \caption{
    Evaluation of algorithm (a) Accuracy w.r.t. ADC resolution without TRQ and (b) with TRQ; (c) Remained A/D operations with TRQ.
  }%
  \label{fig::eval}
  \vspace{-0.4cm}
\end{figure*}






We adopt the ISAAC~\cite{shafiee2016isaac} as our baseline, 
and employ $128 \times 128$ crossbars with single-bit ReRAM cells, 
along with readily available data paths supporting 8b inputs/weights and 16b partial sums.
The parameters of ReRAM and ADC are obtained from~\cite{yao2020fully} and~\cite{chen20183ghz}, respectively.
The system operates at a clock frequency of 100MHz. 
For the digital part, we evaluate our design using data from CACTI6.5 with a 45 nm process for buffers and interconnects, 
The customized peripheral circuits are synthesized with FreePDK 45 nm library\cite{stine2007freepdk} using Design Compiler. 
We build a simulation based on DNN+NeuroSim~\cite{DNN+NEUROSIM}, 
an open-source tool used to evaluate the performance of the neural network on PIM architecture.

To evaluate the accuracy and energy, we use
ResNet-20 on CIFAR-10, ResNet-18, SqueezeNet1.1 on the ImageNet dataset and LeNet-5 on the MNIST dataset. 
We randomly select 32 images from the training dataset as calibration images to fine-tune the ADC configurations. 
The input activations and weights are applied with 8-bit symmetric uniform quantization to accommodate the hardware design. 
Their scaling factors are determined based on the maximum absolute values.
For search space generation, we set $\alpha=0.1$, $\beta=1.2$, the number of range dividing candidates $C=50$, $m$ varies in $[0,7]$. 
This results in up to 400 search candidates.

\subsection{Algorithm Evaluation}
%
%
Fig.~\ref{fig::eval} (a-b) shows the prediction accuracy using different ADC quantization bit-width, 
where ``f/f'' and ``8/f'' represent the model with no quantization (float point)  and 8-bit quantization on \textit{weights} and \textit{activations}, respectively. 
And ``8,7,6,5,4,'' represents the maximum allowed length of \textit{ADC}, i.e., upper bound of $N_{R1}, N_{R2}$\@.
Compared with low-bit uniform quantization, TRQ can achieve better prediction accuracy.
For instance, on ResNet-20 and CIFAR-10 datasets, TRQ achieves 91.09\% prediction accuracy at 4-bit ADC\@, 
while to achieve similar accuracy, U ADC should be at least with 7-bits resolution. 



\subsection{Hardware Evaluation}
\begin{figure}[htbp]
  \centering
  \includegraphics[width=0.46\textwidth]{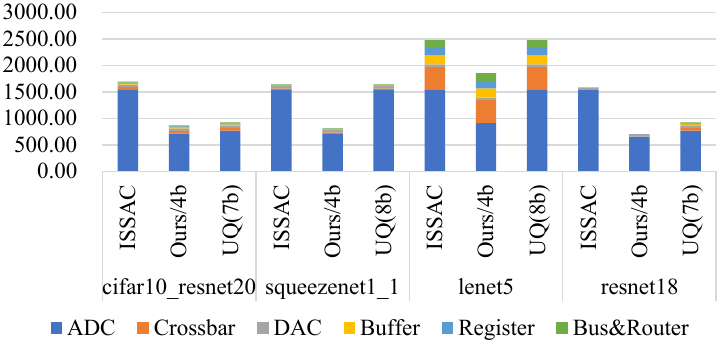}
  \caption{Power breakdown of ReRAM-based accelerator.}%
  \vspace{-0.3cm}
  \label{fig::pbd}
\end{figure} 
In this part, we study how the percentage of ADC dynamic reading efficiency is reduced by TRQ. 
Fig.~\ref{fig::eval} (c) shows the percentage of ADC dynamic reading energy is reduced by TRQ compared with 
A/D operations required by original full precision ADC (8bit/conversion)\@.
The batch size is rescaled for each model across DNNs to keep overall energy in the same range.
The percentage of the ADC energy consumption is reduced to $42\% \sim 62\%$ on average (i.e., $1.6 \sim 2.3 \times$ improvement) with TRQ.
Fig.~\ref{fig::pbd} shows the overall power breakdown of model inference for the above four DNN workloads (the upper bound of $N2$ is set as 4bit). 
We compare 4bit TRQ ADC and U ADC 
that use the minimum bit-width to achieve similar accuracy. 
It can be seen although the original ADC bit-width is unchanged, TRQ can reduce the ADC power consumption significantly.



%% file: tex/6_conclusion.tex
\section{Conclusion}\label{conclusion}


This paper presents an energy-efficient quantization scheme for ReRAM-based neural network accelerators, 
including the TRQ algorithm, hardware implementation, and the algorithm-hardware co-design.
Our design is capable of compressing redundant A/D operations, thereby improving power efficiency with negligible accuracy loss.
Our method can be easily integrated into existing RRAM-based neural network accelerators, requiring no modification of the analog part of the ADC\@, but only a lightweight modification of the digital logic of the ADC\@. 
And the design is transparent to the DNN models, 
requiring no DNN retraining, no overhead for en-/decoding, and \textit{original resolutions} of the ADC is preserved, 
Such flexibility enables our design can adapt to various DNNs and other 
hardware optimizations and model compression techniques without any modification.